\documentclass[conference]{IEEEtran}
\IEEEoverridecommandlockouts
\usepackage[utf8]{inputenc}
\usepackage{amsmath, amssymb}
\usepackage{graphicx}
\usepackage{float}
\usepackage{placeins}
\usepackage{caption}
\usepackage{subcaption}
\usepackage{url}
\usepackage{hyperref}
\usepackage{xcolor}

\usepackage[backend=biber, style=ieee]{biblatex}
\usepackage{braket}
\usepackage{tikz}
\usepackage{float}
\usepackage{soul}
\setstcolor{red}
\usepackage{csquotes}
\usepackage{comment}
\usepackage{algorithmic}
\usepackage{graphicx}
\usepackage{circuitikz}
\usepackage{textcomp}
\usepackage{xcolor}

\begin{document}

\title{Parametric Stability Analysis for Circuit Quantum Electrodynamical Hardwares\\
\thanks{This manuscript has been authored by UT-Battelle, LLC, under Contract No. DE-AC0500OR22725 with the U.S. Department of Energy. The United States Government retains and the publisher, by accepting the article for publication, acknowledges that the United States Government retains a non-exclusive, paid-up, irrevocable, world-wide license to publish or reproduce the published form of this manuscript, or allow others to do so, for the United States Government purposes. The Department of Energy will provide public access to these results of federally sponsored research in accordance with the DOE Public Access Plan. Submitted to the iEE International Conference on Quantum Computing and Engineering - QCE25.}
}

\author{\IEEEauthorblockN{Maria Gabriela Boada}
\IEEEauthorblockA{\textit{Dept. of Physics \& Astronomy} \\
\textit{University of Texas at San Antonio}\\
San Antonio TX, USA \\
mgboada@utsa.edu}
\and
\IEEEauthorblockN{Andrea Delgado}
\IEEEauthorblockA{\textit{Physics Division} \\
\textit{Oak Ridge National Laboratory}\\
Oak Ridge, USA \\
delgadoa@ornl.gov}
\and
\IEEEauthorblockN{Jose Morales Escalante}
\IEEEauthorblockA{\textit{Depts. of Mathematics and Physics \& A.} \\
\textit{University of Texas at San Antonio}\\
San Antonio TX, USA \\
jose.morales4@utsa.edu}
}

\maketitle

\begin{abstract}
The transmon qubit, essential to quantum computation, exhibits disordered dynamics under strong parametric drives critical to its control. We present a combined theoretical and numerical study of stability regions in circuit QED using Floquet theory, focusing on the appearance of Arnold tongues that distinguish stable from unstable regimes. Starting from simple Josephson circuits and progressing to full multimode qubit–cavity systems, we show how time-dependent modulation maps the dynamics to Mathieu-type equations, revealing thresholds for parametric resonances. Perturbative corrections capture effects like higher harmonics and weak nonlinearities. Simulations validate these predictions and expose sensitivity to fabrication parameters. These findings inform thresholds for readout fidelity, amplifier gain, and multi-qubit gate stability.
\end{abstract}


\section{Introduction}

    

Circuit quantum electrodynamics (cQED)~\cite{Blais2004} has emerged as a versatile platform for studying light-matter interactions at the quantum level~\cite{wallraff2004} and for pursuing applications in quantum information processing~\cite{Hofheinz2009, Mirrahimi2014, Heeres2017}. In cQED, superconducting qubits -- artificial atoms with highly engineerable properties -- are coupled to quantized modes of microwave resonators~\cite{BBQ}. 
While cQED devices offer tremendous advantages in terms of fast gate operations and readout~\cite{Pietikainen2024, Gu2021}, experimental work-arounds for improving fidelity and coherence in the prescence of charge noise have come at the cost of increasing hardware complexity~\cite{Somoroff2023, Chou2024, Berdou2023, Wang2016}. Moreover, hardware-efficient strides for next-generation quantum computation have uncovered both the demand and substantial challenges in experimentally realizing complex driven-dissipative processes in cQED~\cite{Leghtas2013, Wang2019, Gertler2021, Li2024}.
For example, strong or multi-tonal parametric pumps can unintentionally excite qubits into highly nonequilibrium states ("ionization")~\cite{Dumas2024,Leghtas2015,Cohen2023Chaos}, revealing the need to understand stability regions in the system's parameter space.
%

In this paper, we explore these instabilities by systematically tracking the emergence of Arnold tongues~\cite{Arnold1965} -- regions in the parameter space of drive frequency and amplitude where solutions to the relevant equations of motion become unbounded. These phenomena  are naturally described by the classical theory of parametric resonances, embodied in the Mathieu equation~\cite{Mathieu1868}. In cQED, a periodically modulated circuit element (for instance, a gate voltage or flux drive) can parameterize the system's effective Hamiltonian in time~\cite{CottetThesis}. By extending the standard Mathieu analysis to include perturbative corrections for damping and quasi-periodicity, we identify thresholds at which the superconducting qubit's unitary oscillations remain bounded or diverge.

We build up our analysis in four stages. First, we study the Cooper pair box (CPB), highlighting its charging energy $E_{C}$ and Josephson energy $E_{J}$ as central parameters. Next, we examine the transmon qubit, which operates in a regime of large $E_{J}/E_{C}$ and enjoys robustness against charge noise at the expense of slightly weaker anharmonicity. We then discuss the electrometer (or Cooper pair transistor), focusing on how the additional junction and measurement bias create a new parametric regime. Finally, we consider the extension of the transistor to the the full multi-mode cQED Hamiltonian, where qubits interact with cavity modes that can also be driven. In each context, the linearization of the circuit dynamics and the subsequent mapping onto Mathieu-type equations reveal how parametric resonances arise from periodic modulations.

Numerical simulations play a crucial role in confirming the analytic predictions, especially in multi-level regimes or when small corrections can significantly shift the boundaries between stable and unstable motion. We chart stability in two-dimensional scans of drive frequency and amplitude, yielding Arnold tongue bifurcations~\cite{Kovacic2018} that capture both primary and higher-order resonances. We incorporate Poincaré sections\cite{Cohen2023Chaos} within the transmon regime to illustrate the sensitivity of parametric instability to device fabrication variations. Finally, we detail the practical ramifications for experiment, how readout power can inadvertently induce qubit ionizations, how parametric amplifiers harness regions of instability deliberately, how multi-qubit gate protocols must avoid resonances, and the natural extension of our stability analysis for the study of open quantum systems.

The black-box quantization method used for mapping to effective Hamiltonians is further detailed in the Appendix.

%
%
%

\section{Equivalency of Tunable Josephson Circuits}

\subsection{Cooper Pair Box}

The CPB is one of the earliest superconducting qubit architectures~\cite{bouchiat1998quantum, nakamura1999coherent, Nakamura1999} and consists of a small superconducting island connected to a reservoir through a single Josephson junction (JJ)~\cite{josephson1962possible}. 
Its energy scales are primarily set by the charge energy 
\begin{equation}
    E_{C}\equiv \frac{e^{2}}{2C_{\sum}}
\end{equation}
(where $C_{\sum}$ is the island's total capacitance) and the Josephson energy $E_{J}$. At zero gate offset charge, the Hamiltonian is

\begin{equation}
    H_{CPB} = 4 E_{C}\hat{N}^{2} - E_{J} cos(\hat{\phi})
    \label{eq: CPB}
\end{equation}

\noindent where $\hat{N}$ is the Cooper pair number operator on the island and $\hat{\phi}$ is the superconducting phase difference across the junction. Introducing a gate offset $N_g$ yields the more general form

\begin{equation}
    H_{CPB} = 4 E_{C}(\hat{N} - N_g)^{2} - E_{J} cos(\hat{\phi})
\end{equation}

In the "charging regime," $E_{C}\gg E_{J}$, the CPB shows strong dependence on $N_g$. Although that confers high charge sensitivity, it also makes the qubit vulnerable to charge noise. From the standpoint of stability analysis, periodically modulating $N_g$ or $E_{J}$ induces time-varying terms in the Hamiltonian that can, near certain biases, be approximated by a Mathieu equation for small deviations about equilibrium.

\subsection{Transmon Qubit}
Although a CPB is highly sensitive to offset charge, the transmon qubit mitigates this charge noise by operating in the large $E_{J}/E_{C}$ regime. In practice, such a device adds a shunt capacitance (or uses a superconducting quantum interference device [SQUID] geometry) to raise $E_{J}$ relative to $E_{C}$. The Hamiltonian still reads

\begin{equation}
    H_{T} = 4 E_{C}(\hat{N} - N_g)^{2} - E_{J}cos(\hat{\phi}),
\end{equation}

but with $E_{J}/E_{C}>>1$. Near the minimum of the cosine potential, one obtains effectively a harmonic oscillator plus a weak nonlinearity, ensuring the first two levels can serve as a qubit with reduced charge dispersion. Nevertheless, strong driving at frequencies commensurate with $2\omega_{01}$ can still trigger parametric instabilities, creating Mathieu-like behaviors. Compared to the CPB, however, these instabilities generally require higher drive amplitudes because of the transmon's reduced anharmonicity.

A split-CPB implements precisely this idea of tuning $E_{J}/E_{C}$ in situ: the CPB junction is split into two junctions in a SQUID loop, so an external flux modulates the effective Josephson energy. With the loop threaded by flux $\Phi$, an offset phase $\delta$ modifies the junction coupling, effectively morphing the device from charge-dominated to transmon-like behavior as $\delta$ shifts the ratio $E_{J}/E_{C}$. In the charge basis, the split-CPB Hamiltonian with junction asymmetry d,
$$
\hat{H}_J = -E_J \frac{1 + d}{2} \cos(\hat{\delta}_1) - E_J \frac{1 - d}{2} \cos(\hat{\delta}_2),
$$
\noindent is entirely equivalent to $H_{CPB}$ with a modified effective josephson energy:

$$E_J^* = E_{J,\Sigma} \cos\left(\frac{\delta}{2}\right) \sqrt{1 + d^2 \tan^2\left(\frac{\delta}{2}\right)}$$

Here, $E_{C}$ is still charging energy, and $E^{*}_{J}$ sets the baseline Josephson coupling. Varying $\delta$ (i.e. flux) continuously adjusts the effective $E_{J}$. Thus, at certain $\delta$-values, the device can look like a standard charge qubit with strong dispersion, whereas at others, it becomes far more transmon-like, with large $E_{J}/E_{C}$ and minimal charge sensitivity.

In order to place the system into realistic parameter range, one can specify gate and junction capacitances, for instance $C_g = 10$ aF and $C_j = 1$ fF. Their sum defines the charging energy via

\begin{equation}
E_C \;=\;\frac{2\,e^2}{2\,\bigl(C_g + C_j\bigr)},
\end{equation}

\noindent while the offset charge $N_g$ derives from a gate voltage $V_g = 2\,e/C_g$. Because the charge basis $\mathbf{n}$ is in principle infinite,  a practical numerical approach truncates $\mathbf{n}$ at some finite window around $\pm (10\,N_g)$. Diagonalizing $\hat{H}_{\text{JJ}}(N_g, \delta)$ for each pair $(N_g, \delta)$ then reveals how energy levels shift as one moves between the "charge" and "transmon" corners of parameter space.

\begin{figure*}[t]
    \centering
    \includegraphics[width=\textwidth]{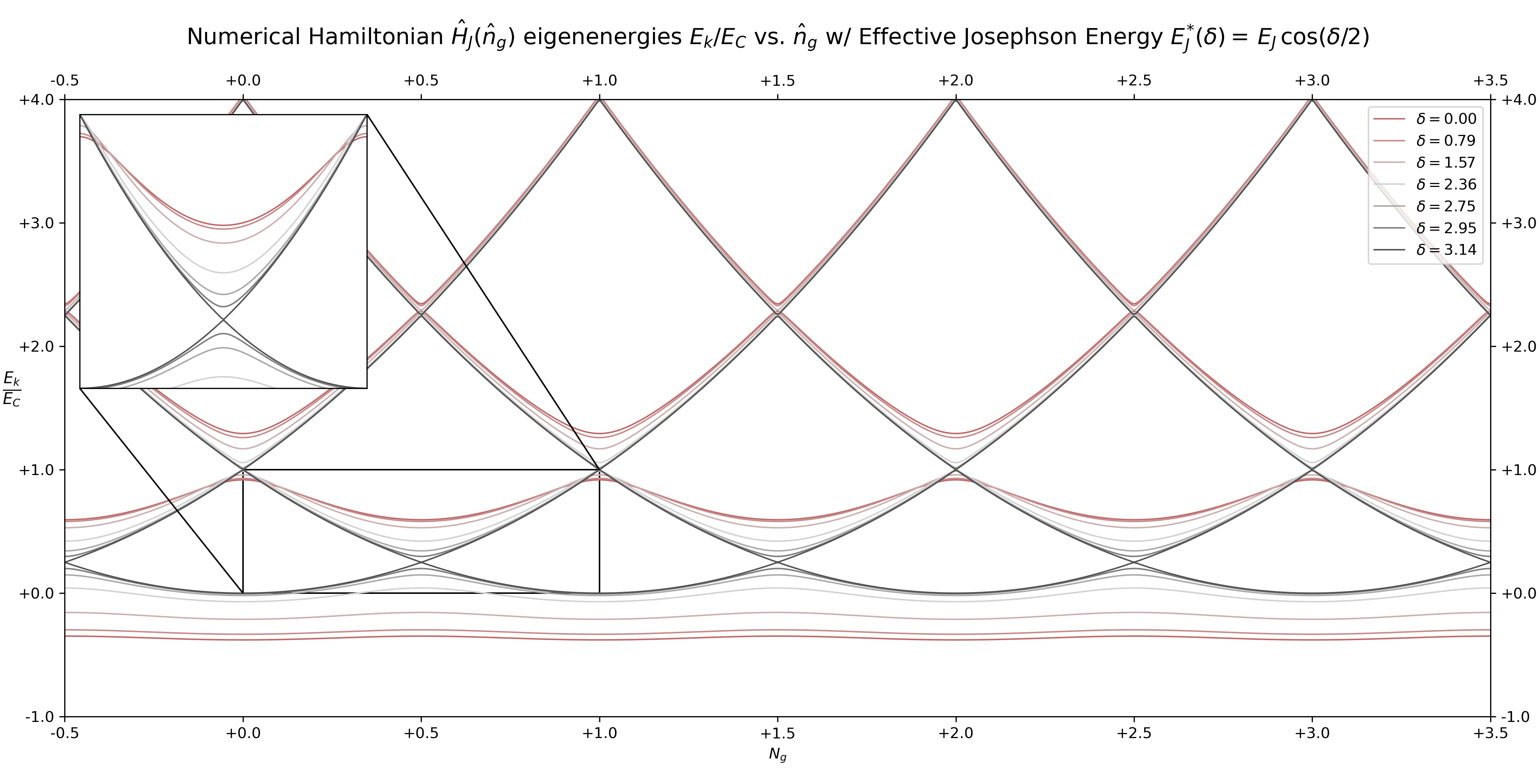}
    \caption{Low-lying eigenenergies of the split Cooper pair box (CPB), 
        plotted as a function of the dimensionless gate charge $N_g$ for 
        several flux (phase) offsets $\delta$. 
        Each colored curve shows the first few energy levels, normalized by 
        the charging energy $E_C$, demonstrating how shifting $\delta$ 
        modifies the effective Josephson coupling and thus reshapes 
        the avoided crossings and overall dispersion.}
        \label{fig:splitCPBenergies}
\end{figure*}

Figure~\ref{fig:splitCPBenergies} illustrates these transitions: plotting the resulting low-lying eigenenergies versus $N_g$ for a range of $\delta$ indicates how flux modifies the avoided crossings or level spacings. When $\delta =0$, one might recover strong Josephson coupling (approaching a transmon limit if $E^{*}_{j}$ is large), whereas $\delta = \pi$ can nearly extinguish the Josephson term, reverting to a more charge-like qubit. Each flux offset yields unique anticrossings, sweet spots, and sensitivities to parametric drives.

Whether one calls the device a “transmon” or “split-CPB” hinges largely on the ratio $E_{J}/E_{C}$ and flux bias. However, the essential physics of parametric driving—particularly at frequencies near $2\omega_{01}$—applies across both ends of this spectrum. A transmon’s weaker nonlinearity means larger drives are required for major instabilities, whereas a CPB with moderate $E_{J}$ can enter Mathieu-like resonance more readily. The split variant then adds the ability to flux-tune that threshold, effectively letting one dial between a charge qubit’s strong parametric response and the transmon’s stability.

Hence, the split-CPB is an ideal platform for studying how parametric effects evolve as $E_{J}/E_{C}$ grows: from robust charge dispersions to a comparatively stable transmon-like regime, all captured within a single device by sweeping $\delta$. This unification of concepts clarifies how transmon qubits are essentially CPBs operating deep in the large $E_{J}/E_{C}$ limit, while the flux tunability of the split-CPB highlights how parametric instabilities can be “turned on” or “turned off” based on the chosen flux bias.

\subsection{Electrometer Configuration}

In the electrometer or Cooper pair transistor (CPT) configuration, a small island is bounded by two junctions in series, forming a device with exceptional charge sensitivity when biased near half-integer $N_g$. Although conceptually similar to the CPB, this arrangement allows strong modulation of the effective Josephson energy with small gate changes, making it an ideal sensor or parametric amplifier element. When driven by an RF or microwave tone, the CPT can exhibit pronounced periodic modulations in its inducive response, again indicating parametric resonance if the drive frequency matches an integer or fractional multiple of the island's natural oscillation frequency.
\subsection{Full cQED System}
The full cQED system introduces a resonator (or cavity) to which one or more superconducting qubits are coupled. A minimal Hamiltonian might be written
\begin{equation}
    H_{cQED} = \hbar \omega_{r}a^{\dagger}a + \sum_{j}E_{j}\ket{j}\bra{j} + \hbar g(a + a^\dagger)\hat{O},
\end{equation}
\noindent where $\omega_{r}$ is the resonator frequency, $a^{\dagger},a$ are the creation and anihilation operators for photons, and $\hat{O}$ is related to the qubit variables (e.g., charge or phase). In the dispersive limit, one can reduce this to an effective model where qubit excitations shift the resonator frequency, but intense driving can go beyond this simplification, especially when the drive is near or at parametric resonance conditions. The interplay of cavity damping, the transmon's weak anharmonicity, and higher-level states further complicates the stability analysis, often favoring numerical methods to confirm predictions of parametric instabilities.

\section{Lumped-Circuit/Black-Box Quantization Approach}
A useful and highly effective framework for analyzing superconducting qubit circuits is the \emph{lumped-element} or \emph{black-box approach} introduced in~\cite{BBQ}, often referred to as the BBQ method. In this viewpoint, one considers the qubit-containing element—such as a CPB or transmon—in combination with its surrounding environment as a single “black box.” By measuring or simulating the circuit’s linear response, one identifies dressed harmonic modes (including parasitic capacitances, inductances, and resonator modes) and extracts their effective admittances or impedances.

Replacing the intricate geometry and multi-mode structure with a succinct lumped model of parallel and series LC oscillators allows one to incorporate the Josephson junction as an additional \emph{nonlinear} element, characterized by its junction inductance $L_{J}$ and the purely nonlinear portion of the Josephson potential. One can then write an effective Hamiltonian that accounts for each harmonic mode—associated with its resonance frequency $\omega_m$—plus perturbative quartic (or higher-order) corrections from the junction. In the transmon regime, where $E_J \gtrsim E_C$, this leads to a reduced description:

\begin{equation}
\hat{H} = 4E_C(\hat{n} + \hat{n}_r)^2 - E_J \cos \hat{\phi} - \sum_m \hbar\,\omega_m \hat{a}_m^\dagger \hat{a}_m
\end{equation}

capturing both the qubit’s anharmonic levels and the linear modes of the environment. By systematically retaining only the most relevant modes or strongest couplings, one obtains a tractable few-mode Hamiltonian that still reflects the true complexity of the underlying circuit.

From the standpoint of \emph{parametric instability}, this lumped-circuit viewpoint clarifies how a periodic drive—applied to one or more dressed modes—can yield parametric amplification or bifurcation when the drive frequency is near twice (or another rational fraction of) one of these modes. The Josephson junction’s nonlinearity, once expressed in terms of these collective coordinates, generates interaction terms capable of triggering multiphoton processes or forming Arnold tongues under strong drive conditions. Adopting or combining such a black-box reduction with Floquet theory (which systematically treats time-periodic Hamiltonians) offers a powerful predictive tool for designing and operating cQED devices in regimes of strong drive and significant nonlinearity.

While earlier sections introduced the qubit Hamiltonian in terms of $\hat{N}$ and $\hat{\phi}$ directly, the black-box method affords a more direct route for complex, multi-mode devices. It makes it easier to see how certain parametric drives couple primarily to specific modes, and to identify the frequencies at which the system might become unstable. The resulting lumped-element model can greatly streamline calculations, particularly when many stray inductances and capacitances would otherwise obscure the essential qubit–mode interactions.

\begin{figure}[h]
    \centering
    \includegraphics[width=\linewidth]{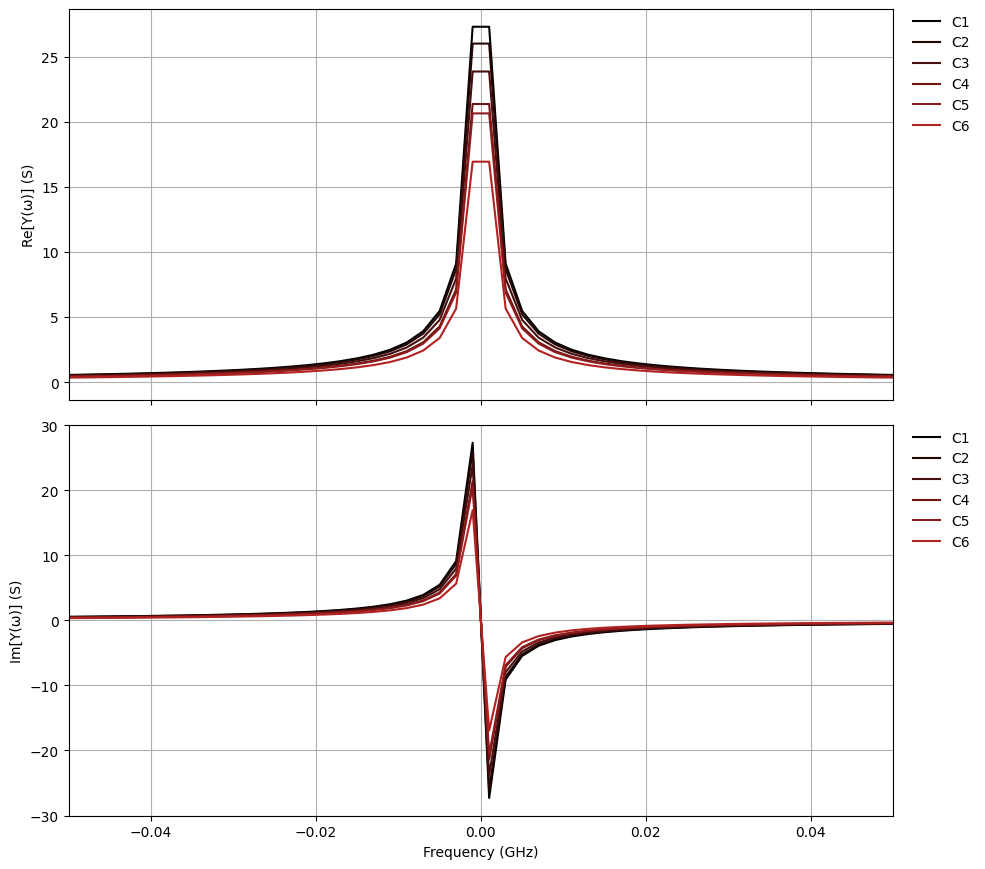}
    \caption{Real and imaginary parts of the linear admittance \( Y(\omega) \) for six representative circuits used in black-box superconducting circuit quantization, labeled C1 through C6. The plots show the frequency dependence near resonance, with slight variation in peak shape and width across the different configurations.}
    \label{fig:BBQadmittance}
\end{figure}

The cornerstone of black‑box quantization is the \emph{linear} impedance
$Z(\omega)$ (equivalently the admittance $Y(\omega)=Z^{-1}(\omega)$) that the
non‑linear element sees once the surrounding electromagnetic environment has
been lumped into effective components. For the multi-mode networks sketched in Fig.~\ref{fig:BBQadmittance}, the impedance can be expressed in the compact
form

\begin{flalign}
    Z(\omega)=\sum_{j=1}^{N_j}
    \Bigl[
        i\omega C_j^{*}
        +\frac{1}{i\omega L_j^{*}}
        +\frac{1}{R_j^{*}}
    \Bigr]^{-1}
    \;=\;
    Y^{-1}(\omega),
\label{eq:Zsum}
\end{flalign}

\noindent where the starred parameters $C_j^{*},L_j^{*},R_j^{*}$ encode geometric and material details for each resonant branch \cite{BBQ}. Setting $R_j^{*}\!\to\!\infty$ discards dissipation and leaves a set of loss‑less LC–oscillators whose poles at $\omega=\omega_j$ determine the normal
modes.

Near every pole one introduces flux amplitudes $f_j$ which, after canonical
quantisation, become
\begin{equation}
    f^{\phantom{\dagger}}_j,\;f_j^{\dagger}\;\longrightarrow\;
    \sqrt{\frac{\hbar}{2}\,\mathcal Z^{\text{eff}}_j}\,a^{\phantom{\dagger}}_j,
    \quad
    \sqrt{\frac{\hbar}{2}\,\mathcal Z^{\text{eff}}_j}\,a_j^{\dagger},
\end{equation}
with
\begin{equation}
    \mathcal Z^{\text{eff}}_j=
    \frac{2}{\omega_j\,\operatorname{Im} Y'(\omega_j)}.
\label{eq:Zeff}
\end{equation}
Here $a_j$ and $a_j^{\dagger}$ annihilate and create single quanta in the
$j$‑th mode, while $\mathcal Z^{\text{eff}}_j$ plays the role of an effective
impedance extracted directly from the derivative of the admittance.

Using~\eqref{eq:Zeff}, the total branch flux and its conjugate charge expand as
\begin{align}
    \hat{\phi} &= \sum_{j=1}^{N_j}\hat{\phi}_j,
    &
    \hat{\phi}_j &= 
        \sqrt{\frac{\hbar}{2}\,Z_j^{\text{eff}}}\,
        \bigl(a^{\phantom{\dagger}}_j + a_j^{\dagger}\bigr),
\label{eq:phi_modes}
\\[2pt]
    \hat{Q} &= -i\sum_{j=1}^{N_j}\hat{Q}_j,
    &
    \hat{Q}_j &= 
        \sqrt{\frac{\hbar}{2}\,\frac{1}{Z_j^{\text{eff}}}}\,
        \bigl(a^{\phantom{\dagger}}_j - a_j^{\dagger}\bigr).
\label{eq:Q_modes}
\end{align}
The ladder operators themselves follow the canonical form
\begin{align}
    a^{\phantom{\dagger}}_j &=
        \sqrt{\frac{1}{2\hbar Z_j^{\text{eff}}}}\,
        \hat{\phi}_j
        +
        i\sqrt{\frac{Z_j^{\text{eff}}}{2\hbar}}\,
        \frac{\hat{Q}_j}{\omega_j},
\\[2pt]
    a_j^{\dagger} &=
        \sqrt{\frac{1}{2\hbar Z_j^{\text{eff}}}}\,
        \hat{\phi}_j
        -
        i\sqrt{\frac{Z_j^{\text{eff}}}{2\hbar}}\,
        \frac{\hat{Q}_j}{\omega_j},
\end{align}
so that each pole of $Y(\omega)$ is promoted to an independent quantum harmonic
oscillator.

Replacing $(a_j,a^{\dagger}_j)$ by the mode expansions
(\ref{eq:phi_modes})–(\ref{eq:Q_modes}) in the linearised circuit plus the full
Josephson non‑linearity leads to
\begin{equation}
    \hat{H}_{\text{nd}}=
    \sum_{j}\hbar\bar{\omega}_j\,a_j^{\dagger}a^{\phantom{\dagger}}_j
    \;+\;
    E_J\!\left[\cos\hat{\phi}+\frac{\hat{\phi}^{2}}{2}\right],
\label{eq:H_blackbox}
\end{equation}
with $\hat{\phi}=\hat{\phi}/\Phi_0$ and $\Phi_0=\hbar/2e$.  Equation
\eqref{eq:H_blackbox} embeds all linear electromagnetic degrees of freedom in
the ladder operators while retaining the full Josephson cosine as the unique
source of anharmonicity.  When a time‑periodic flux modulation drives one of
the modes, the reduced dynamics of $\hat{\phi}$ can be mapped onto a
Mathieu‑type oscillator (see next Section), whose instability tongues will be analyzed in
Sec.~\ref{sec:ArnoldTongues}. 

Together, the CPB, transmon, electrometer, and full cQED models form a hierarchy of devices that progressively incorporate more complexity and physical realism. This structured approach allows us to trace how parametric instabilities emerge and evolve across different regimes of circuit operation. In the sections that follow, we use this device hierarchy to frame our analysis of Mathieu-type dynamics and Floquet-based stability, providing both generalized insights and device-specific predictions. See Appendix for derivation of the reduced CPB Hamiltonian from the black-box formulation.

\section{Mapping to a Mathieu Equation and Perturbative Corrections}

In what follows, we describe a general recipe for mapping Hamiltonians to Mathieu equations. 
A central idea uniting these diverse cQED elements is that once a superconducting circuit is driven by a time-dependent parameter -- whether it be a gate voltage, flux, or cavity drive -- its effective equation of motion may approximate a Mathieu equation. The classical Mathieu equation is 

\begin{equation}
    \frac{d^{2}x}{dt^{2}} + (\delta + \epsilon cos \Omega t)x = 0
\label{eq:mathieu}
\end{equation}

\noindent and is known to exhibit stability lobes in the $(\delta, \epsilon)$ plane. Quantum mechanically, one can study the same phenomenon by examining how a periodically modulated Hamiltonian leads to exponential growth or bounded oscillations in the state vector.

In transmon, for instance, one might consider small oscillations of $\hat{\phi}(t)$ obeys an equation with a time-varying coefficient near $\omega_{p} \equiv 8E_{J}E_{C}/\hbar^{2}$, then a resonance condition $\Omega\approx 2\omega_{p}$ can induce parametric amplification of $\hat{\phi}(t)$. Even in more elaborate setups, such as a cavity drive that couples indirectly to the transmon, the rotating-wave approximation can fail under strong driving, thereby admitting a Floquet analysis of the full Hamiltonian that reproduces key features of the Mathieu equation.

The essence of this generalized picture is that parametric resonance emerges when the drive's frequency and amplitude align such that energy is pumped into the system's oscillatory degrees of freedom faster than it can be lost to damping or redistributed among other modes. In quantum terms, this typically appears as multi-photon excitations that push the qubit from $\ket{0}$ to higher levels or populate a cavity mode to high photon numbers.

Even as we map Eq.~\ref{eq: CPB} at its most generic circuit parameters, 

\begin{equation}
\frac{d^2\!f}{dt^2} + \left[\,4\frac{E_k}{E_C} + \frac{E_J}{E_C} \cos (2t)\right]\,f(t) = 0
\label{eq: MathieuGeneric}
\end{equation}

\noindent the flux-tunable Josephson circuit cast to Mathieu form bears a significant amount of physicality.  For example, Eq.~\ref{eq: MathieuGeneric} is $\pi-$periodic on $t$ (phase) \cite{CottetThesis}, where the gate charge 
\[f_k(t)= \exp(-iN_g)\,x_k(t)\]
is absorbed into the role of the characteristic exponent due to the circular topology associated with charge states \cite{CottetThesis}.
For an approximate treatment of the energy levels for $N_g\in[0,1]$ where $k(m,\, N_g)$ is defined as the sum over $\ell\pm1$ over $k_\ell(m, N_g)$ where 
\begin{align}
k_\ell(m, N_g) = 
&\left[\, \mathrm{int}(2N_g + \ell / 2) \bmod 2 \,\right] \nonumber 
\times \\
&\left\{ \mathrm{int}(N_g) + \ell(-1)^m \left[ (m+1) / 2 \right] \right\},
\end{align}
is chosen such that for each value of $N_g$, only a discrete set of values of $k(m, N_g)$ are possible \cite{CottetThesis}. Thus, the characteristics 
for the respective even and odd Mathieu functions of order
n, given the parameter
$\epsilon$, $a_n
 (\epsilon)$, $b_n
 (\epsilon)$, can be used to write the approximate analytical expression for an $E_k$ for arbitrary $N_g$ given $k(m,\,N_g)$.

\begin{figure}[h]
    \centering
    \includegraphics[width=\linewidth]{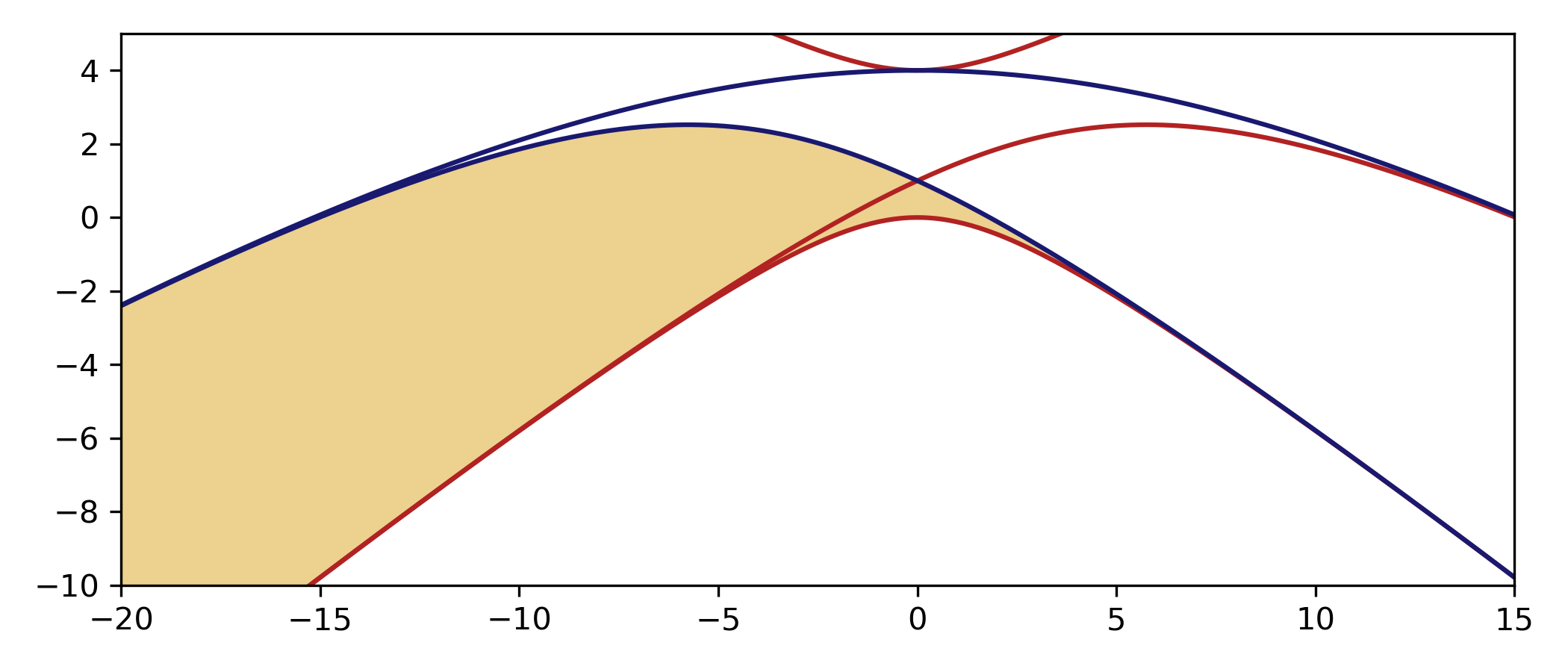}
    \caption{Bifurcation diagram illustrating Arnold tongue structures arising from the Mathieu equation. The shaded regions represent zones of parametric stability, enclosed by the intersections of even and odd characteristic curves (shown in blue and red, respectively). These boundaries correspond to resonance conditions where the characteristic exponents become real and bounded. Stability arises in regions where the even and odd solutions alternate, producing bands in which the Floquet solutions remain bounded. Outside these regions, the solutions grow exponentially, indicating instability.}
    \label{fig:arnold_mathieu}
\end{figure}

\section{Derivation of Stability Regions (Arnold Tongues)}

Having established the theoretical mapping to Mathieu equations for various Josephson-based circuits, we now apply Floquet theory to identify stability regions in the drive-parameter space. These regions, or Arnold tongues, directly relate to the device architectures discussed in Section II, particularly in understanding how different regimes of EJ/EC impact susceptibility to parametric excitation.

Floquet theory provides a natural way to identify stability regions in periodically driven systems. One looks for solutions of the form $\Psi(t) = e^{\mu t}p(t)$, where $p(t)$ is a time-periodic function. If the real part of the Floquet exponent $\mu$ is positive, the solution grows unboundedly, marking instability. If $\mu$ is purely imaginary, oscillations remain bounded. Tracing out curves where $Re(\mu) = 0$ defines the boundaries in drive-amplitude-drive-frequency space that separate stable from unstable motion.

In the simplest classical Mathieu equation, one typically encounters a series of "tongues" or "lobes" emanating from resonance conditions $\Omega = 2\omega_{p}/m$ (Fig.~\ref{fig:arnold_mathieu}). The largest (primary) tongue appears near the fundamental parametric resonance $\Omega\approx2\omega_{p}$, with smaller ones at integer or fractional multiples. Once damping and higher-order corrections are included, these tongues shift and shrink, but they retain the essential lob-like topology. Physically, each lobe represents a region where the system can synchronously absorb energy from the drive.

In cQED contexts, especially for the transmon, the principal instability zone often arises near $\Omega \approx 2\omega_{01}$. Secondary resonances may appear around $\Omega \approx \omega_{01}$, $\Omega\approx 2\omega_{01}/3$, or other rational fractions, but are generally narrower. Because the transmon has weaker anharmonicity than the CPB, it also tends to exhibit narrower tongues that require larger drive amplitudes to become manifest. By contrast, a strongly charge-sensitive CPB can display quite pronounced tongues, indicating it is more easily driven into instability under moderate parametric excitation.

Identifying these tongues is crucial for experimental design. Operating within an unstable region can lead to uncontrolled qubit excitations and severly limit the device's functionality. Conversely, parametric amplifiers intentionally exploit such unstable regions to achieve large gain, so precisely locating the boundary can guide amplifier performance.

\section{Numerical Simulations}
To validate the analytic insights derived in Sections II and IV, we conduct numerical simulations of the Mathieu equation, focusing on parameter ranges relevant to CPB and transmon devices. These simulations bridge the simplified models of Section II with real-world scenarios, including the effects of damping and fabrication-induced variability.

While many cQED systems require a full quantum treatment—typically involving integration of a time-dependent Schrödinger or master equation—certain regimes of parametric driving can be captured accurately by classical equations of motion. The Mathieu equation is one such classical model, describing a pendulum with time-varying stiffness. Its solutions can become unbounded (unstable) when the drive frequency and amplitude satisfy resonant conditions, forming the well-known Arnold tongue structure in parameter space.

To illustrate this, one can perform direct numerical integration of the classical Mathieu equation (Eq.~\ref{eq:mathieu}), where $\delta$ and $\epsilon$ control the natural frequency and the strength of the parametric modulation, respectively, and $\omega$ is the drive frequency. By choosing an initial displacement $x(0)$ and velocity $\dot{x}(0)$ and numerically integrating over a sufficiently long duration, one can determine whether the solution remains bounded (Stable) or grows beyond some threshold (unstable).

{In the}
\emph{undamped} version of the Mathieu equation, the second derivative of $x$ depends only on the instantaneous value of $x$. 
{The Mathieu oscillator with a
\emph{damping term} $\gamma\dot{x}$, representing energy losses, is} such that the equation becomes

\begin{equation}
    \frac{d^{2}x}{dt^{2}} + \gamma\frac{dx}{dt} + (\delta + \epsilon cos \Omega t)x = 0.
\label{eq:mathieudamp}
\end{equation}

Damping generally suppresses the area of instability lobes, so the tongues in the parametric space $(\delta, \epsilon)$ are narrower compared to the undamped case.

To generate an Arnold tongue diagram, one sweeps through a grid of $\delta$ and $\epsilon$ values (with fixed $\omega$ and $\gamma$), integrates the Mathieu equation over time, and uses a simple criterion to classify the final outcome as “stable” or “unstable.” A typical heuristic is to check if the maximum amplitude exceeds a chosen threshold by the end of the simulation; if it remains below the threshold, we mark it stable. Plotting stability (e.g., 1 for stable and 0 for unstable) against $\delta$ and $\epsilon$ then yields a two-dimensional colormap whose lobes indicate parametric resonance.

\begin{figure}[htbp]
\centering
\includegraphics[width=\linewidth]{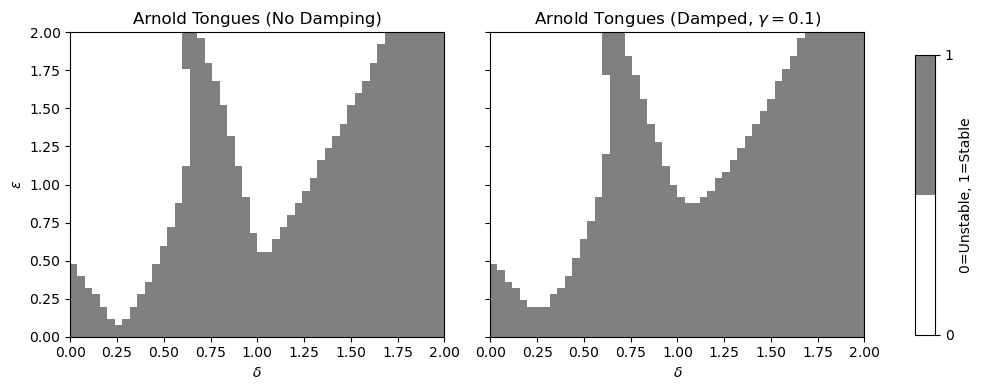}
\caption{Side-by-side numerical Arnold tongue diagrams for the Mathieu equation with and without damping. The left panel shows the undamped system, where parametric instabilities form wide lobes as a function of $\delta$ and $\epsilon$. The right panel includes a damping term $\gamma = 0.1$, which diminishes and shifts the unstable regions. The colormap uses two shades: white for “unstable” (0) and gray for “stable” (1).}
\label{fig:mathieutongue}
\end{figure}

An example is shown in Fig.~\ref{fig:mathieutongue}. The results are displayed side by side -- one chart for $\gamma=0$ (undamped) and another for a finit $\gamma$, illustrating how dissipation shrinks or shifts the unstable zones.

Although the classical Mathieu approach does not capture all quantum aspects of circuit QED (e.g., discrete transmon levels, multi-photon transitions, or master-equation effects), it does provide a useful first approximation of parametric instabilities. In scenarios where the qubit or oscillator is well described by a single degree of freedom and moderate drives (within the validity of linearization), these classical simulations highlight precisely where large-amplitude motion is expected—akin to where a quantum system might exhibit strong excitations or “ionization” out of its potential well. For a more complete quantum treatment, one could in principle replace the classical integration with time-dependent Schrödinger or master-equation simulations, but for analyzing archetypal Arnold tongue phenomena, the classical Mathieu equation remains a powerful and computationally efficient tool.

\section{Results}
\label{sec:results}

\subsection{Bifurcation and Arnold Tongue Plots}
\label{sec:ArnoldTongues}

A central part of our analysis involves identifying where the split or conventional CPB undergoes parametric instability under periodic driving, a phenomenon often visualized through bifurcation or Arnold tongue diagrams. In the absence of damping, these diagrams are particularly stark, revealing extensive lobes of unstable solutions once the drive amplitude and frequency coincide with resonances of the qubit. While real devices inevitably include dissipation, the no-damping picture highlights the fundamental thresholds beyond which small perturbations would push the system into large-amplitude motion.

\begin{figure}[h]
    \centering
    \includegraphics[width=\linewidth]{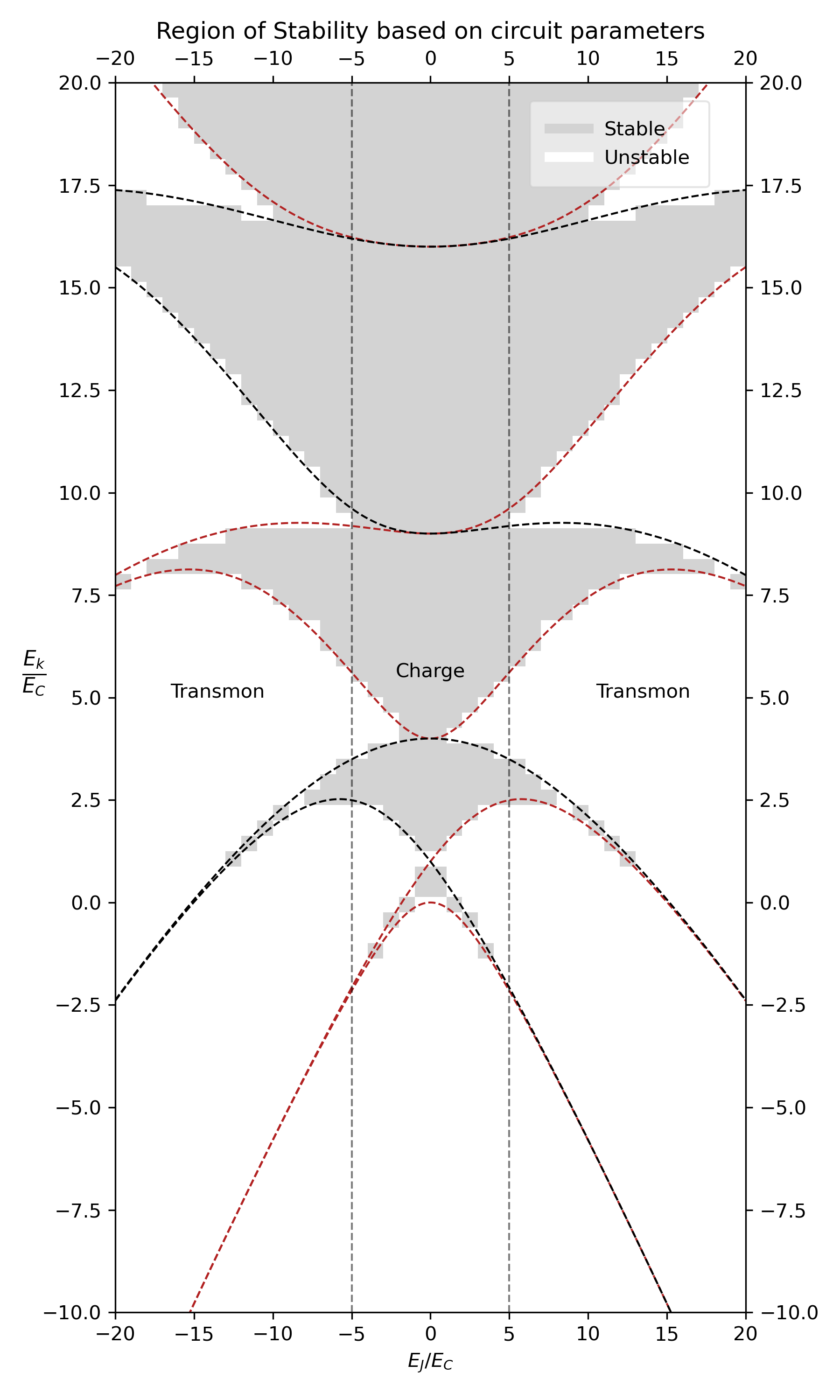}
    \caption{The Mathieu equation describes parametric resonance and has the associated stable and unstable zones, as indicated above. Transitions between stability regions occur where characteristic curves $a_n(q)$ and $b_n(q)$ intersect or bifurcate. These transitions define the boundaries of ``Arnold tongues'' in the stability diagram and correspond to changes in the nature of solutions from bounded to unbounded.}
    \label{fig:bifurcation_diagram}
\end{figure}

Fig.~\ref{fig:bifurcation_diagram} plots a vertical axis of $E_{k}/E_{C}$, corresponding to low-energy levels (or parameterized energies) normalized by the charging energy $E_{C}$. The horizontal axis is $E_{J}/E_{C}$, the ratio of Josephson to charging that typically distinguishes "charge" from "transmon" operating regimes. In the center, around $E_{J}/E_{C}\approx0$, the device is dominated by charging effects, hence one sees larger dispersions or avoided crossings reminescent of the pure CPB regime. Far to the right (and left), where $|E_{J}/E_{C}|$ becomes large, the system moves towards a flux- or transmon-like regime, characterized by flatter level spacings and reduced sensitivity to offset charge.

The shaded gray regions in the figure designate "stable" parameter zones under the chosen drive conditions, while the white pockets (or tongues) represent "unstable" parameter ranges. Thus, if the system is biased at an ($E_{J}/E_{C}, E_{k}/E_{C}$) point lying within a white region, a small drive at or near the corresponding resonance can cause significant amplitude growth --- an indication of parametric excitation or bifurcation. By contrast, remaining in the gray region means that similar drives will not lead to unbounded oscillations, making it a safer choice for typical qubit operations.

Dashed lines in black and red trace approximate boundaries or theoretical expansions for these instability zones (e.g., from semiclassical expansions around the potential minima), and the shaded regions represent numerically determined stability zones, with the boundaries confirmed through detailed integration of the system dynamics. The precise edges of these regions are better visualized as continuous stability contours that validate theoretical predictions. One typically observes symmetric (or nearly symmetric) lobes about the central charge regime, indicating that when $\pm|E_{J}/E_{C}|$ is large, the device resides in an effectively transmon-like or flux-tuned domain, and parametric resonances shift accordingly.

Overall, this bifurcation diagram underscores the inherent nonlinear thresholds faced by a tunable JJ as one sweeps from the charge to the transmon regime. It reinforces the central theme: circuit parameters such as $E_{J}/E_{C}$ drastically affect whether a given device is safe from parametric excitation or primed for strong nonlinear amplification. In future work, we will incorporate damping and more realistic conditions to refine these boundaries, yet the present figure solidifies the fundamental mechanisms of parametric bifurcation in the superconducting qubit architecture.

\begin{figure*}[t]
    \centering
    \includegraphics[width=\linewidth]{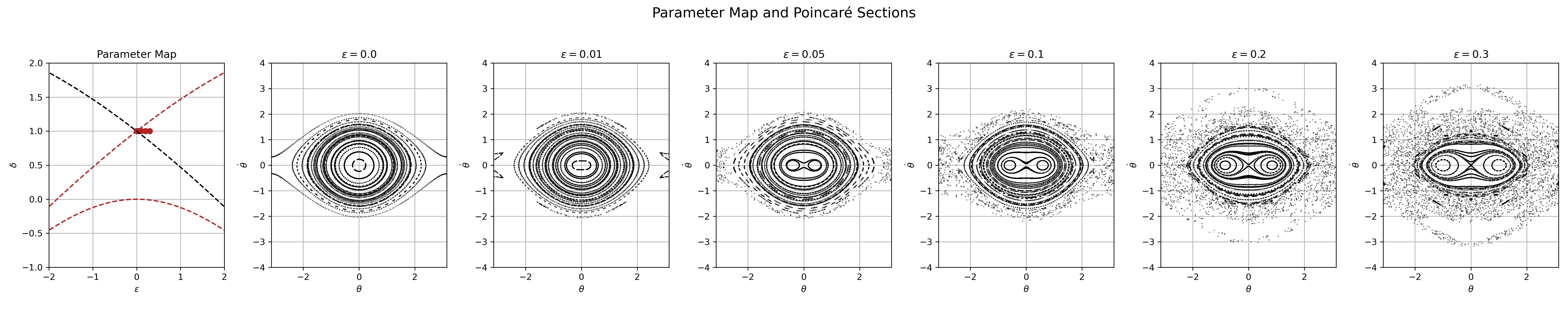}
    \caption{Parameter map (left) and Poincaré sections (right) illustrating the onset of chaos in a periodically driven pendulum as the drive amplitude \( \epsilon \) increases. The red point on the parameter map marks the fixed detuning \( \delta \) and varying \( \epsilon \) values used for each section. For small \( \epsilon \), the dynamics are quasiperiodic with invariant tori dominating phase space. As \( \epsilon \) increases, these tori are progressively broken, giving rise to resonance islands and stochastic layers. At higher drive strengths, widespread torus destruction leads to global chaos, characterized by irregular, ergodic trajectories filling large regions of phase space.}
    \label{fig:poincare-pendulum}
\end{figure*}

\subsection{Poincaré Sections and the Onset of Chaos}

Poincaré sections provide a compact visual diagnostic of order versus chaos in nonlinear dynamics. In a conventional phase portrait, the Poincare section is a cross-section drawn at some time T in the $(x, \dot{x})$ plane. In a stroboscopic sampling, one takes the trajectory only at discrete times $t=kT$ ($k \in \mathbb{N}$), where $T$ is the period of external drive or some integer multiple of it. For the periodically driven oscillators considered here the underlying equation of motion is a differential equation in which a single degree of freedom is modulated by an external perturbation $\theta$. Plotting the stitched-together points $(x(t=kT), \dot{x(t=kT)})$ yields a stroboscopic snapshot of the flow.

The geometry of the resulting point cloud conveys immediate information" an \emph{eye-shaped} structure (a separatrix) signals a robust, coherent oscillation that persists under moderate perturbation; an \emph{infinity-shaped} curve betrays a region of mixed regular and irregular motion; a \emph{uniform spray} of dots indicates fully developed chaos, where nearby initial conditions diverge rapidly and no simple periodicity survives. A separatrix partially covered by scattered points foreshadows sensitive dependence on initial conditions: even minor perturbations can push the system into the disordered sea.

Fig.~\ref{fig:poincare-pendulum} juxtaposes a parameter map (left) with representative Poincaré sections (right panels) for a parametrically driven pendulum. The fixed detuning $\delta$ is marked by a red dot, while the driving amplitude $\epsilon$ is increased along the ordinate. For small $\epsilon$ the Poincaree section consists of thin, intact closed curves. These are intersections of \emph{KAM tori} --- the quasi-periodic invariant surfaces guaranteed by the Kolmogorov-Arnold-Moser theorem for a nearly integrable Hamiltonian --- with the stroboscopic plane. As $\epsilon$ increases, these tori are gradually torn apart: first narrow resonance islands appear, then broad stochastic layers, and finally the closed loops disintegrate into a sea of scattered points, signaling the onset of global chaos.

\begin{figure*}[t]
    \centering
    \includegraphics[width=\textwidth, height=.375\textwidth]{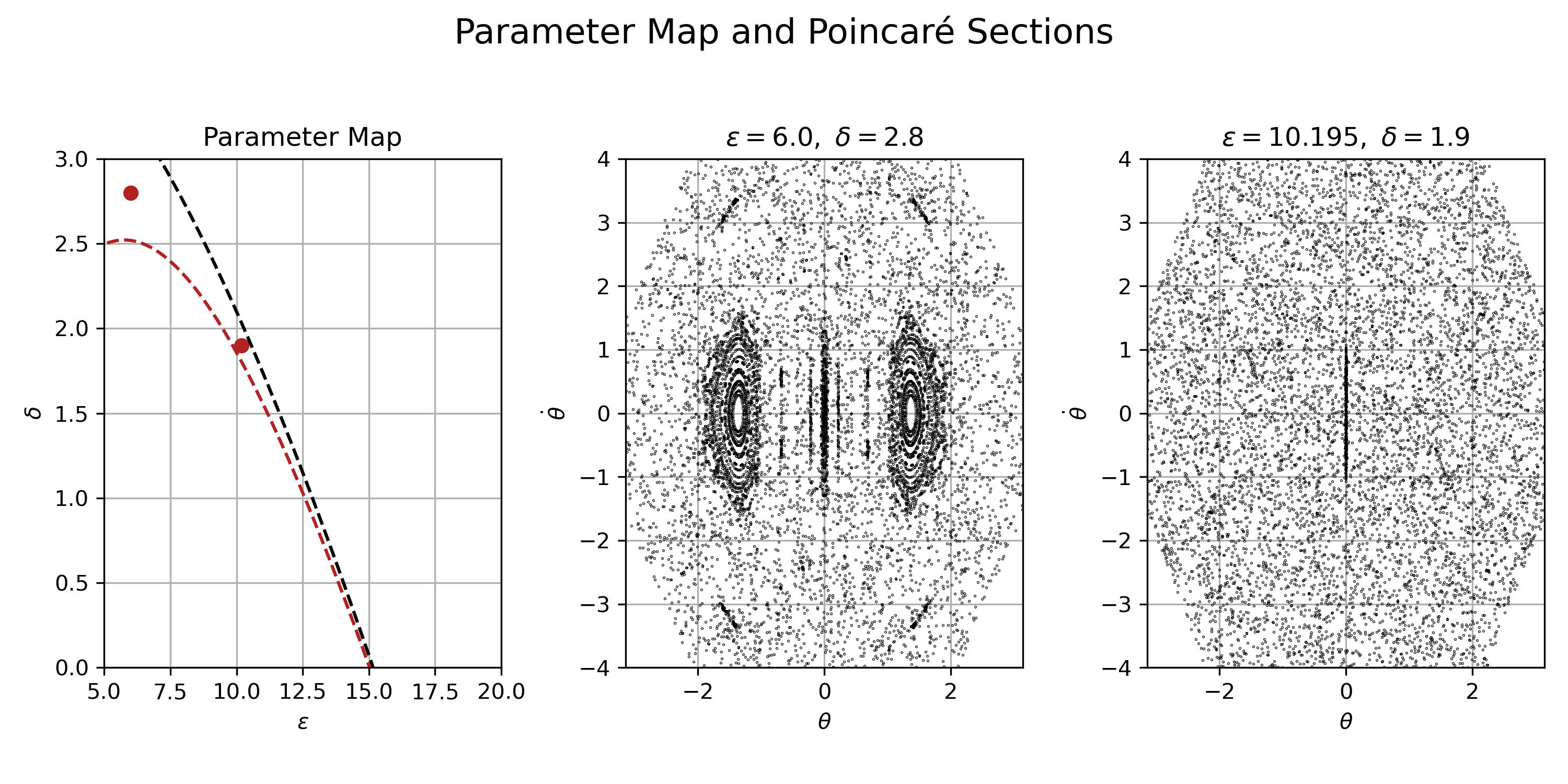}
    \caption{Poincaré section for a split-CPB biased in the transmon regime. Although linear analysis suggests bounded oscillations, a small perturbation of the initial conditions produces a broad scatter of points, pointing to disordered behavior and underscoring thee device sensitivity to drive in this parameter range.}
    \label{fig:poincare-transmon}
\end{figure*}

An analogous study for the split CPB ( or its transmon limit) is shown in Fig.~\ref{fig:poincare-transmon}. Here, the system is biased deep in the transmon regime, where linear theory predicts bounded motion. The Poincaré section, however, reveals that even a modest perturbation can seed a widespread irregularity, demonstrating that the apparently "safe" operating point is precariously close to chaotic dynamics.

These two examples reinforce the broader narrative of this paper: linear stability analysis and Arnold tongue boundaries indicate \emph{where} parametric instabilities begin, but Poincaré sections reveal \emph{how} the ensuing nonlinear dynamics evolve whether the system remains confined to smooth resonance islands or is swept into global chaos that compromises qubit fidelity. Mapping these sections across the $(\delta, \epsilon)$ plane therefore provides a practical tool for locating safe operating zones and for engineering controlled, yet stable, parametric interactions in cQED hardware.

\section{Experimental Relevance}

The emergence of Arnold tongues in cQED has direct implications for device operation and experimental design. One compelling example is the trade-off between readout fidelity and qubit stability. Fast, high-fidelity readout often demands a strong measurement drive on the cavity or on the qubit itself, but pushing this drive too far may place the system in or near an instability lobe, causing the qubit to leave its low-energy manifold (an effect sometimes referred to as “transmon ionization”). As soon as the qubit escapes the well—or equivalently, is driven to high excitation numbers—its transition frequency shifts or it saturates, thereby degrading measurement contrast. By consulting the Arnold tongue diagrams, experimentalists can locate a drive amplitude below the threshold of instability while still achieving sufficient signal-to-noise.

In a related vein, parametric amplifiers and bifurcation-based detectors deliberately operate in or near instability regions to achieve large gain or bistability. Their principle of operation relies on driving a nonlinear system into a regime where small input signals can tip it between two states of oscillation. Knowing the precise shape and threshold of these tongues informs both the design (e.g., the ratio of inductances and capacitances that sets $\omega_{p}$) and the choice of operating points (drive frequency and amplitude) for optimum performance.

In multi-qubit cQED systems, strong time-dependent pulses can inadvertently drive parametric resonances if the pulses contain frequency components near rational multiples of qubit transitions. This can lead to leakage errors or correlated errors across multiple qubits. By determining where these resonances lie, one can engineer pulses that selectively avoid or exploit them. Moreover, future cQED devices that incorporate multiple modes or qubits will have a higher-dimensional set of potential resonances, making a thorough analysis of stability an increasingly essential design tool.

\section{Conclusion and Future Directions}

We have demonstrated how Floquet theory and the concept of Arnold tongues illuminate the behavior of superconducting qubits under strong parametric drives. By stepping through key cQED elements—the Cooper pair box, the transmon, the electrometer configuration, and the full qubit–cavity system—we have shown that time-dependent modulations naturally lead to Mathieu-like instabilities. The classic lobes from the Mathieu equation persist, although damping, higher harmonics, and multi-level corrections complicate the details. Numerical simulations confirm these theoretical predictions, while Monte Carlo variations in device parameters illustrate how real-world fabrication tolerances can shift a qubit in and out of potential instability regimes.

The experimental relevance of these findings is already seen in phenomena such as transmon ionization during high-power readout and the controlled use of parametric amplifiers. More sophisticated multi-qubit circuits, advanced gate schemes, and novel approaches to readout all underscore the need to chart these stability boundaries carefully. In future work, one may investigate how quantum fluctuations or finite temperature modify the onset of instability, or how to harness rapid transitions across these tongues for qubit reset or nonclassical state generation. Moreover, as device sizes grow, interactions between many qubits and multiple cavity modes can yield more complex dynamical phase diagrams, but the underlying Floquet-based approach remains a flexible way to identify and control parametric instabilities.

By providing both an analytic and numerical framework, as well as exploring how design variations affect stability, our analysis equips cQED engineers with insights needed to maximize coherence and control fidelity while mitigating unwanted transitions. 
{The interplay of differential equation methods
such as Floquet theory and, on the other hand,
superconducting circuit design,} thus continues to shape the next generation of robust, high-performance quantum devices.

\section*{Acknowledgment}
This material is based upon work supported by the U.S. Department of Energy, Office of Science, Office of Workforce Development for Teachers and Scientists, Office of Science Graduate Student Research (SCGSR) program. The SCGSR program is administered by the Oak Ridge Institute for Science and Education for the DOE under contract number DE‐SC0014664. This work was partially supported by the U.S. Department of Energy, Office of Science, Office of Nuclear Physics Quantum Horizons: QIS Research and Innovation for Nuclear Science program at ORNL under FWP ERKBP91.

\printbibliography

\appendix
\section*{Mapping the Black-Box Hamiltonian Back to the Cooper Pair Box}

Starting from the transmon Hamiltonian of Ref.~\cite{Blais2020}\,\,(Eq.\,(20) therein),)

\begin{equation}
\hat{H}_T = 4E_C(\hat{n} - N_g)^2 - E_J \cos \hat{\phi},
\qquad
\hat{n} = \hat{Q}/2e, 
\qquad
\hat{\phi} = \frac{2\pi\hat{\Phi}}{\Phi_0}
\end{equation}

with $E_C = e^2/2C_\Sigma$ and $\Phi_{0}=h/2e$, we outline how the black-box formulation reduces to the familiar CPB form once charge coupling to an external LC mode is reinstated.

For $E_J / E_C \gg 1$ the offset charge $N_g$ hardly perturbed the spectrum, so we set $N_g\!=\!0$:

\begin{equation}
\hat{H}_T = 4E_C \hat{n}^2 - E_J \cos \hat{\phi}
\label{eq:H_noNg}
\end{equation}

Attaching the island to a resonator through a small gate capacitance $C_{g}$ adds a second dynamical node (labeled $C$). Writing the Lagrangian and moving to Hamiltonian form one obtains

\begin{equation}
\hat{H}= 
\frac{1}{2C_{\Sigma}}\!
\Bigl(\hat{Q}_{A}+\tfrac{C_{g}}{C}\hat{Q}_{C}\Bigr)^{2}
- E_{J}\cos\hat{\phi}_{A}
\;+\;
\underbrace{\frac{\hat{Q}_{C}^{2}}{2C}+\frac{\hat{\Phi}_{C}^{2}}{2L}}_{\displaystyle
\hat{H}_{\mathrm{LC}}},
\end{equation}

where $\hat{Q}_A$ is the island charge and $C_{\Sigma}=C_{g}+C_{J}$.

Introducing $\hat{n}_r = (C_g / C)\hat{Q}_C / 2e$ and $\hat{n} = \hat{Q}_A / 2e$, yields

\begin{equation}
\hat{H} = 4E_C(\hat{n} + \hat{n}_r)^2 - E_J \cos \hat{\phi}_A + \hat{H}_{LC}
\label{eq:H_shiftedCharge}
\end{equation}

The resonator's bare Hamiltonian is then diagonalized in terms of normal-mode operators:

\begin{equation}
\hat{H}_{\mathrm{LC}}
=
\sum_{m}\hbar\omega_{m}\, \hat{a}_{m}^{\dagger}\hat{a}_{m},
\qquad
\hat{n}_{r}=\sum_{m}\hat{n}_{m},
\quad
\hat{n}_{m}= \frac{C_{g}}{C_{m}}\frac{\hat{Q}_{m}}{2e}.
\end{equation}

Combining Eqs.~\ref{eq:H_shiftedCharge} and the mode expansion one obtains the compact form quoted on the main text,

\begin{equation}
\boxed{\;
\hat{H}=4E_{C}\bigl(\hat{n}+\hat{n}_{r}\bigr)^{2}
- E_{J}\cos\hat{\phi}
+ \sum_{m}\hbar\omega_{m}\,\hat{a}_{m}^{\dagger}\hat{a}_{m}\; }.
\end{equation}

Hence, the bare CPB couples capacitively to each normal mode via shifted charge $\hat{n}_{r}$. In a one-mode truncation ($\omega_{r}$, capacitance $C_{r}$) the Hamiltonian reduces to the textbook cQED form~\cite{Koch2007}:

\begin{equation}
\hat{H}=4E_{C}(\hat{n}-N_g)^{2}
- E_{J}\cos\hat{\phi}
+ \hbar\omega_{r}\,\hat{a}^{\dagger}\hat{a} 
+ 2\beta e V_{\mathrm{rms}}^{0}\,\hat{n}\,(\hat{a}+\hat{a}^{\dagger}),
\end{equation}
where $\beta=C_{g}/C_{\Sigma}$ and
$V_{\mathrm{rms}}^{0}=\sqrt{\hbar\omega_{r}/2C_{r}}$. Any residual offset charge $N_g$ omitted in Eq.\,\eqref{eq:H_noNg} can be
re‑introduced perturbatively when $E_{J}/E_{C}$ is large but finite.




\end{document}